\documentclass[onecollarge]{svjour2}       
\smartqed  

\usepackage[dvipdfmx]{graphicx}

\usepackage{amsmath, amssymb}
\usepackage{amsfonts}
\usepackage{multicol}
\usepackage{braket}

\usepackage{ascmac}
\usepackage{fancybox}
\usepackage{comment}

\usepackage{cite}

\newcommand{\eq}[1]{\begin{equation} #1 \end{equation}}
\newcommand{\eqa}[2]{\begin{equation} #1 \label{#2} \end{equation}}
\newcommand{\balign}[1]{\begin{align} #1 \end{align}}

\newcommand{\todayd}{\the\year/\the\month/\the\day}
\newcommand{\del}{\partial}
\newcommand{\bib}{\bibitem}

\newcommand{\lb}{\label}
\newcommand{\nt}{\notag}

\newcommand{\ft}[2]{\left. #1 \right|_{#2}}

\newcommand{\eref}[1]{Eq.~\eqref{#1}}

\newcommand{\sref}[1]{Sec.\ref{s:#1}}
\newcommand{\cref}[1]{Chap.~\ref{c:#1}}

\newcommand{\bel}{\begin{easylist}}
\newcommand{\eel}{\end{easylist}}

\def \({\left(}
\def \){\right)}
\newcommand{\la}{\left\langle}
\newcommand{\ra}{\right\rangle}
\def \[{\left[}
\def \]{\right]}

\newcommand{\sumtwo}[2]%
{\mathop{\sum_{#1}}_{#2}}
\newcommand{\sumthree}[3]%
{\mathop{\mathop{\sum_{#1}}_{#2}}_{#3}}
\newcommand{\sumfour}[4]%
{\mathop{\mathop{\mathop{\sum_{#1}}_{#2}}_{#3}}_{#4}} 
\newcommand{\prodtwo}[2]%
{\mathop{\prod_{#1}}_{#2}}
\newcommand{\mintwo}[2]%
{\mathop{\min_{#1}}_{#2}}
\newcommand{\maxtwo}[2]%
{\mathop{\max_{#1}}_{#2}}
\newcommand{\maxthree}[3]%
{\mathop{\mathop{\max_{#1}}_{#2}}_{#3}}
\newcommand{\limtwo}[2]%
{\mathop{\lim_{#1}}_{#2}}
\newcommand{\suptwo}[2]%
{\mathop{\sup_{#1}}_{#2}}
\newcommand{\supthree}[3]%
{\mathop{\mathop{\sup_{#1}}_{#2}}_{#3}}
\newcommand{\supfour}[4]%
{\mathop{\mathop{\mathop{\sup_{#1}}_{#2}}_{#3}}_{#4}} 
\newcommand{\inftwo}[2]%
{\mathop{\inf_{#1}}_{#2}}
\newcommand{\infthree}[3]%
{\mathop{\mathop{\inf_{#1}}_{#2}}_{#3}}
\newcommand{\inffour}[4]%
{\mathop{\mathop{\mathop{\inf_{#1}}_{#2}}_{#3}}_{#4}} 
\newcommand\calA{{\cal A}}

\newcommand\calF{{\cal F}}

\newcommand\calI{{\cal I}}
\newcommand\calJ{{\cal J}}

\newcommand{\bsp}{\boldsymbol{p}}

\newcommand{\ep}{\varepsilon}

\newcommand{\hJ}{\hat{J}}

\newcommand{\hX}{\hat{X}}

\newcommand{\dsgm}{\dot{\sigma}}

\newcommand{\pss}{p^{\rm ss}}
\newcommand{\hZ}{\hat{Z}}

\newcommand{\Var}{{\rm Var}}
\newcommand{\hcalJ}{\hat{\calJ}}
\newcommand{\hcalI}{\hat{\calI}}
\newcommand{\hn}{\hat{n}}
\newcommand{\dPi}{\dot{\Pi}}

\def\rnum#1{\resizebox{0.5em}{\height}{\expandafter{\romannumeral #1}}}
\def\Rnum#1{\resizebox{0.5em}{\height}{\uppercase\expandafter{\romannumeral #1}}}

\begin{document}

\title{Optimal thermodynamic uncertainty relation in Markov jump processes}


\author{Naoto Shiraishi}


\institute{Naoto Shiraishi \at Department of Physics, Gakushuin University, 1-5-1 Mejiro, Toshima-ku, Tokyo 171-8588, Japan \\
             \email{naoto.shiraishi@gakushuin.ac.jp}           
          }

\date{Received: date / Accepted: date}

\maketitle

\begin{abstract}

We investigate the tightness and optimality of thermodynamic-uncertainty-relation (TUR)-type inequalities from two aspects, the choice of the Fisher information and the class of possible observables.
We show that there exists the best choice of the Fisher information, given by the pseudo entropy production, and all other TUR-type inequalities in a certain class can be reproduced by this tightest inequality.
We also demonstrate that if we observe not only generalized currents but generalized empirical measures, the TUR-type inequality becomes optimal in the sense that it achieves its equality in general nonequilibrium stationary systems.
Combining these results, we can draw a hierarchical structure of TUR-type inequalities.

\end{abstract}

\section{Introduction}\label{intro}

In the last decade, the thermodynamic uncertainty relation (TUR) has attracted much attention in the field of nonequilibrium statistical mechanics, particularly in stochastic thermodynamics~\cite{Sei12, Shi21}.
The TUR connects two different quantities, the fluctuation of a current around a nonequilibrium stationary state and the entropy production, in systems with time-reversal symmetry.
The TUR was conjectured by Barato and Seifert~\cite{BS15}, and first proved by Gingrich and co-authors~\cite{Ging16, GRH17, HG17} by using elaborated large deviation techniques.
Various studies~\cite{PBS16, PBS16-full, PLE16, FPS18, Bar19} settle along this line with a large deviation approach, while the computation is highly complicated and a transparent prospect was still veiled.
The refinement of the proof of the TUR was supplied by a series of papers of Dechant and Sasa~\cite{DS18, DS20, DS21} and Dechant~\cite{Dec19}, where the TUR can be understood as a particular case of the generalized Cram\'{e}r-Rao inequality with a proper perturbation on the path probability.
From this viewpoint, the entropy production is related to the Fisher information, and the averaged current is understood as a consequence of the rescaling of time-scale.
This highly simplifies the proof of the TUR, which allows us to extend and generalize the TUR easily.

The extension of the TUR has been discussed in various directions:
One direction is to relax some assumptions on the setup of the TUR and generalize the inequality to a wider class of systems.
Examples include systems with arbitrary initial states~\cite{LGU20}, systems with time-dependent driving~\cite{KS20}, discrete-time processes~\cite{Shi17-note, PB17, LGU20}, systems without time-reversal symmetry~\cite{BHS17, MBG18, Bar19} including underdamped Langevin systems~\cite{FPS18, LPP19, LPP21, KKL21}, waiting-time statistics~\cite{GH17, Pig17, Gar17, FE20, HS21}, and processes with unidirectional transitions~\cite{PRR21}.
Another direction is to replace the entropy production with other quantities such as activity~\cite{Gar17, TB19, HS21} and probability current~\cite{PRR21}, which provide a new type of TUR-type inequalities.
The short-time limit of the TUR~\cite{PRS17, DS18-PRE, Li19, Ots20} has also been discussed in the context of inference of entropy production~\cite{Li19, Ots20}. 
Note that preceding results of this type of inequalities can be seen in various contexts including power-efficiency trade-off relation in heat engines~\cite{SST16, SS19} and speed limit inequalities~\cite{Ito18, SFS18, VVH20}.

Throughout these studies, what is {\it good} or {\it ultimate} forms of the TUR is an important question from fundamentals to applications.
Revealing the most fundamental form of the TUR will enhance our understanding of the structure in stochastic thermodynamics, and a tighter bound will be more helpful for applications.
However, unfortunately, the original TUR is observed not to achieve its equality in general nonequilibrium conditions~\cite{Ots20, HH18}, which suggests some missing gap.
This fact motivates several studies to modify the TUR and find tighter bounds which are hoped to achieve their equalities~\cite{DS21, DS21-2, KS21, Kam21}.
Although these studies shed new light on the TUR, the full picture of the optimality of TUR-type inequalities has still not been addressed.

In this paper, we clarify the relationship between various TUR-type inequalities and particularly elucidate their tightness from a unified viewpoint.
We tune two quantities, the Fisher information and the class of observables, to reach the optimal form of TUR-type inequalities.
In the first part, we optimize the Fisher information and find the tightest TUR-type inequality as long as we adopt a general and standard method to derive TUR-type inequalities.
Various other TUR-type inequalities can be derived from the tightest one.
In the second part, we expand the class of possible observables from only generalized currents to the sum of generalized currents and generalized empirical measures, and establish the optimality of the TUR, where the equality is achieved in general nonequilibrium stationary systems.
As its corollary, we find that if we observe only generalized currents, the TUR-type inequality cannot achieve its equality.

This paper is organized as follows.
In \sref{review}, we provide a quick derivation of the general form of TUR-type inequalities based on the latest short proof.
This serves as a pedagogical introduction to the TUR and TUR-type inequalities.
In \sref{Fisher}, we prove that the TUR-type inequality with the {\it pseudo entropy production} is the tightest as long as we employ a standard method for TUR-type inequalities.
In \sref{observable}, we extend the class of observables to generalized currents and empirical measures and demonstrate the optimality of the TUR-type inequality.

\section{Quick derivation of thermodynamic uncertainty relation with the generalized Cram\'{e}r-Rao inequality}\lb{s:review}
\subsection{Thermodynamic uncertainty relation and its general framework}

We consider a continuous-time Markov jump process on discrete states in $0\leq t\leq \tau$.
The time evolution of the probability distribution is given by the following master equation:
\eq{
\frac{d}{dt}p_i(t)=\sum_j R_{ij}p_j(t),
}
where $p_i(t)$ is the probability distribution of the state $i$ at time $t$, and $R$ is the transition matrix. 
The transition matrix $R$ satisfies the nonnegativity of off-diagonal elements ($R_{ij}\geq 0$ for $i\neq j$) and the normalization condition ($\sum_i R_{ij}=0$).
Throughout this paper, we suppose that the transition matrix is time-independent.

We introduce a time-integrated generalized current defined as
\eq{
\hcalJ _d :=\sum_{(i,j)} d_{ij} \hcalJ_{ij}.
}
Here, $\hcalJ_{ij}:=\hn_{ij}-\hn_{ji}$ is the conventional probability empirical current, where $\hn_{ij}$ is the number of jumps from the state $j$ to the state $i$, and $d_{ij}$ is an antisymmetric weight: $d_{ij}=-d_{ji}$.
The ensemble average of this stochastic process is denoted by the bracket $\la \cdot \ra$, and the quantity without a hat represents its ensemble average (e.g., $\calJ_d=\langle \hcalJ_d\rangle$).
The instant generalized current at time $t$ is given by
\eq{
J_d(t)=\sum_{i\neq j}d_{ij}R_{ij}p_j(t).
}
Quantities with superscript ``ss" represent those in the stationary distribution $\pss$.

We suppose the {\it local detailed-balance condition}, with which the entropy production rate $\dsgm$ is given by
\eqa{
\dsgm (t)= \sum_{i,j} R_{ij}p_j(t)\ln \frac{R_{ij}p_j(t)}{R_{ji}p_i(t)} .
}{LDB}
The thermodynamic uncertainty relation (TUR) claims that if the system is in the stationary distribution, the following inequality holds:
\eqa{
\frac{(\calJ_d^{\rm ss})^2}{\Var(\calJ_d)}\leq \frac{\sigma}{2},
}{TUR}
where $\sigma:=\int_0^\tau dt \dsgm(t)$ is the entropy production and $\Var(\calJ_d):=\langle (\hcalJ_d -\langle \hcalJ_d\rangle)^2\rangle$ is the variance of $\hcalJ_d$.

The TUR can be extended to a system whose initial distribution is not necessarily the stationary distribution.
Other conditions are the same as the conventional TUR.
In this setup, Liu, Gong, and Ueda~\cite{LGU20} showed the following TUR-type inequality:
\eqa{
\frac{(\tau J_d(\tau))^2}{\Var(\calJ_d)}\leq \frac{\sigma}{2}.
}{GTUR1}
In stationary systems, \eref{GTUR1} reduces to \eref{TUR} since $\tau J_d(\tau)=\calJ_d^{\rm ss}$.

\bigskip

To derive these inequalities, we first prove a general framework for TUR-type inequalities.
To this end, we shall show
\eqa{
\frac{(\calJ_d^{\rm ss})^2}{\Var(\calJ_d)}\leq \frac{\calF}{2}
}{FTUR-ss}
for stationary systems and 
\eqa{
\frac{(\tau J_d(\tau))^2}{\Var(\calJ_d)}\leq \frac{\calF}{2}
}{FTUR}
for systems with arbitrary initial states, with
\eq{
\calF:=\int_0^\tau dt \sum_{i\neq j}F_{ij}(t).
}
We can change $F$ in a certain class of functions which is specified later.
By changing $F$, we can derive several TUR-type inequalities.
With this respect, \eref{FTUR-ss} and \eref{FTUR} can be regarded as a general form of TUR-type inequalities.

\subsection{Proof of the general form of TUR}

To prove Eqs.~\eqref{FTUR-ss} and \eqref{FTUR}, we introduce the {\it generalized Cram\'{e}r-Rao inequality}.
The generalized Cram\'{e}r-Rao inequality treats the situation that the probability distribution of a probabilistic variable $x$ depends on a parameter $\theta$, and our task is to estimate the value of $f(\theta)$ from the variable $x$ by using a function $g(x)$.
Let $\la F(x) \ra_\theta:=\int dx F(x)P_\theta(x)$ be the average with the probability distribution $P_\theta(x)$.
The Fisher information of the probability distribution parametrized by $\theta$ is defined as~\cite{CTbook}
\eq{
F(\theta):=-\la \frac{\del^2}{\del \theta^2}\ln P_\theta(x)\ra_\theta =\la \( \frac{\del}{\del \theta}\ln P_\theta(x)\) ^2\ra_\theta .
}
Suppose that $g(x)$ is an unbiased estimator of $f(\theta)$ (i.e., $\la g(x)\ra_\theta=f(\theta)$).
Then, the generalized Cram\'{e}r-Rao inequality states that the variance of the unbiased estimator $g(x)$ denoted by ${\rm Var}_\theta (g):=\la (g-\la g\ra_\theta)^2\ra_\theta=\la (g-f(\theta))^2\ra_\theta$ satisfies
\eqa{
{\rm Var}_\theta (g)\geq \frac{(f')^2}{F(\theta)}.
}{CR}

The generalized Cram\'{e}r-Rao inequality \eqref{CR} is a direct consequence of the Schwarz inequality:
\balign{
{\rm Var}_\theta (g) F(\theta)=&\( \int dx (g(x)-f(\theta))^2 P_\theta(x)\) \( \int dx \( \frac{\del}{\del \theta}\ln P_\theta(x)\) ^2 P_\theta(x)\) \nt \\
\geq& \( \int dx (g(x)-f(\theta)) \( \frac{\del}{\del \theta} \ln P_\theta(x) \) P_\theta(x) \) ^2 \nt \\
=& \( \frac{\del }{\del \theta} \int dx g(x) P_\theta(x)  \) ^2 \nt \\
=&(f')^2.
}
Here, in the fourth line, we used $\frac{\del }{\del \theta}\int dx P_\theta(x)=0$, which follows from $\int dx P_\theta(x)=1$ for all $\theta$.

We now apply the generalized Cram\'{e}r-Rao inequality \eqref{CR} to Markov jump processes.
We set $x$ as a stochastic trajectory $\Gamma$, $P_\theta$ as the path probability, and $g(x)$ as the generalized current $\hcalJ_d$.
This particular choice of \eref{CR} is named {\it fluctuation-response inequality} by Dechant and Sasa~\cite{DS20}.
We employ the parametrized path probability $P_\theta(\Gamma)$ with the following transition matrix
\balign{
R^\theta_{ij}&=R_{ij}e^{\theta Z_{ij}}, \lb{Rtheta-1} \\
R^\theta_{jj}&=-\sum_i R_{ij}e^{ \theta Z_{ij}}, \lb{Rtheta-2}
}
and set $\theta=0$.
Note that $Z_{ij}$ is time-dependent in general.
If we write $\Var$ and $\langle \cdot \rangle$ without subscript $\theta$, this means that these quantities are at $\theta=0$.
In this choice, the Fisher information reads
\balign{
F(0)=&\int_0^\tau dt \[ \sum_{i\neq j} R_{ij}p_j(t) \frac{\del^2}{\del \theta^2} \ln R^\theta _{ij}+\sum_j p_j(t) \frac{\del^2}{\del \theta^2} \ln e^{-R^\theta _{jj}} \] \nt \\
=&\int_0^\tau dt \sum_{j} \[ p_j(t)  \frac{\del^2}{\del \theta^2} \sum_{k(\neq j)}R_{kj}e^{ \theta Z_{kj}} \] \nt \\
=&\int_0^\tau dt \sum_{k\neq j} R_{kj}p_j(t)Z_{kj}^2 \nt \\
=&\la \int_0^\tau dt \hZ^2 \ra ,  \lb{FRI-KL-Z2}
}
where we used $\frac{\del^2}{\del \theta^2} \ln R^\theta _{ij}=\frac{\del^2}{\del \theta^2} \theta Z_{ij}=0$ for $i\neq j$ in the second line.
Thus the generalized Cram\'{e}r-Rao inequality is written as
\eqa{
\( \frac{\del \langle \hX\rangle_\theta}{\del \theta} \)^2\leq {\rm Var} (X) \int_0^\tau dt \langle \hZ^2\rangle 
}{FRI-Z2}
for any observable $X(\Gamma)$, which will be set to a generalized current $\calJ_d$ in our case.

\bigskip

Suppose that for all $i$ and $j$ we have $Z_{ij}(t)$ satisfying the following two conditions:
\balign{
(Z_{ij}(t))^2R_{ij}p_j(t)+(Z_{ji}(t))^2R_{ji}p_i(t)=&F_{ij}(t), \lb{Zcond-1} \\
Z_{ij}(t)R_{ij}p_j(t)-Z_{ji}(t)R_{ji}p_i(t)=&R_{ij}p_j(t)-R_{ji}p_i(t), \lb{Zcond-2}
}
with some function $F_{ij}(t)$.
The first condition \eqref{Zcond-1} is connected to the left-hand side of Eqs.~\eqref{FTUR-ss} and \eqref{FTUR}.
The second condition \eqref{Zcond-2} is responsible for rescaling the time-scale:
 The current in $\theta$ system is $(1+\theta)$-fold of that in the reference system with $\theta=0$ up to $O(\theta)$:
\balign{
J_{ij}^{ \theta}(t)=&R^\theta _{ij}p_j(t)-R^\theta _{ji}p_i(t) \nt \\
=&R_{ij}p_j(t)-R_{ji}p_i(t)+\theta(Z_{ij}(t)R_{ij}p_j(t)-Z_{ji}(t)R_{ji}p_i(t))+O(\theta^2) \nt \\
=&(1+\theta)J_{ij}(t)+O(\theta^2). \lb{theta-current}
}

We first consider the case of the stationary system, which corresponds to \eref{FTUR-ss}.
In the stationary state, \eref{theta-current} suggests that the stationary distribution of $\theta$-system is the same as the reference system with $\theta=0$ up to the order $O(\theta)$ (i.e., $\bsp^{\theta, {\rm ss}}=\bsp^{\rm ss}+O(\theta^2)$) and all the stationary currents grow $(1+\theta)$-fold, which directly implies
\eq{
\ft{\frac{\del \langle \hcalJ_d \rangle_\theta}{\del \theta}}{\theta=0}=\calJ_d.
}
Combining \eref{Zcond-1}, we arrive at the desired relation \eqref{FTUR-ss}.

\bigskip

We next consider the case with arbitrary initial states, which corresponds to \eref{FTUR}.
In this case, the second condition for $Z$, \eref{Zcond-2}, yields
\balign{
R^\theta \bsp(t)=&(1+\theta) R\bsp(t) +O(\theta^2). 
}
From this relation, we assert that the probability distribution driven by $R^\theta$ is given by $\bsp' (t)=\bsp((1+\theta)t)+O(\theta^2)$ (time length is rescaled by $(1+\theta)$-fold), which is confirmed as
\balign{
\frac{d}{dt}\bsp'(t)=&(1+\theta)\ft{\frac{d}{dt'}\bsp(t')}{t'=(1+\theta)t} \nt \\
=&(1+\theta)\ft{\frac{d}{dt'}\bsp(t')}{t'=t}+\ft{\frac{d}{dt''}\( \ft{\frac{d}{dt'}\bsp(t')}{t'=t''}\)}{t''=t} \cdot \theta t+O(\theta^2) \nt \\
=&R^\theta \bsp(t)+\ft{\frac{d}{dt'}\( R^\theta \bsp(t')\)}{t'=t} \cdot \theta t+O(\theta^2) \nt \\
=&R^\theta \bsp((1+\theta)t)+O(\theta^2) \nt \\
=&R^\theta \bsp'(t)+O(\theta^2) .
}
Here we used $R-R^\theta =O(\theta)$ in the third line.
Thus we regard the $\theta$-system as a system with rescaling of timescale $t\to (1+\theta)t$, which enables us to compute
\eqa{
\ft{\frac{\del \langle \hcalJ_d \rangle_\theta}{\del \theta}}{\theta=0}= \ft{\frac{\del}{\del \theta} \int_0^{(1+\theta)\tau} dt J_d(t)}{\theta=0}=\tau J_d(\tau)
}{GTUR1-J}
Plugging \eqref{GTUR1-J} into \eqref{FRI-Z2}, we arrive at the desired result, \eref{FTUR}.

\subsection{Proof of TUR}

Observe that the choice
\eqa{
F_{ij}(t)=\dot{\Pi}_{ij}(t):=\frac{(R_{ij}p_j(t)-R_{ji}p_i(t))^2}{R_{ij}p_j(t)+R_{ji}p_i(t)}
}{Fij-choice}
has the following solution of $Z_{ij}(t)$ satisfying Eqs.~\eqref{Zcond-1} and \eqref{Zcond-2}:
\eqa{
Z_{ij}(t)=\frac{R_{ij}p_j(t)-R_{ji}p_i(t)}{R_{ij}p_j(t)+R_{ji}p_i(t)}.
}{Z-choice}
We call the quantity $\Pi$ as {\it pseudo entropy production}, whose importance in the context of TUR has been mentioned~\cite{Dec-private}.
Using an elementary inequality $2(a-b)^2/(a+b)\leq (a-b)\ln a/b$ for $a,b>0$, we readily have
\eqa{
\Pi :=\int_0^\tau dt \sum_{i\neq j}\dot{\Pi}_{ij}(t)\leq \int_0^\tau dt \sum_{i\neq j}R_{ij}p_j(t)\ln \frac{R_{ij}p_j(t)}{R_{ji}p_i(t)} =\sigma .
}{Pi-sigma}
Inserting \eref{Pi-sigma} into \eref{FTUR-ss} and \eref{FTUR}, we arrive at the TUR \eqref{TUR} for stationary systems and \eref{GTUR1} with arbitrary initial states.

\bigskip

We put a remark on Eqs.~\eqref{FTUR-ss} and \eqref{FTUR}.
In the derivation of these inequalities, we have not used the local detailed-balance condition \eqref{LDB}.
The local detailed-balance condition is used to connect the pseudo entropy production $\Pi$ to the entropy production $\sigma$.

\section{Optimal Fisher information}\lb{s:Fisher}

\subsection{Psuedo entropy production as the best parameter}

We have derived a general form of TUR, \eqref{FTUR-ss} and \eqref{FTUR}, and these inequalities provide various TUR-type inequalities including the original TUR \eqref{TUR} by setting $F$ and $Z$ properly.
Previous investigations of TUR-type inequalities can be regarded as the investigation of $Z$, good parametrizations of the path probability.
From this standpoint, it is natural to ask what choice of $F$ and $Z$ is a good choice to obtain a tight inequality.
In this section, we shall demonstrate that the TUR-type inequality with $F=\dot{\Pi}$, which has been shown previously~\cite{Dec-private} and we have already seen in the previous section, is in fact the tightest TUR-type inequality as long as we follow the derivation based on the generalized Cram\'{e}r-rao inequality with conditions \eqref{Zcond-1} and \eqref{Zcond-2}.
In other words, 
\eqa{
\frac{(\calJ_d^{\rm ss})^2}{\Var(\calJ_d)}\leq \frac{\Pi}{2}
}{master-TUR-ss}
for stationary systems and 
\eqa{
\frac{(\tau J_d(\tau))^2}{\Var(\calJ_d)}\leq \frac{\Pi}{2}
}{master-TUR}
with arbitrary initial states are the tightest TUR-type inequality serving as the {\it master} inequality, and other TUR-type inequalities with other $F$ and $Z$ are derived by just applying an elementary inequality
\eq{
\Pi\leq \calF.
}
Namely, we no longer need to seek a proper parametrization $Z$ in the path probability.

To demonstrate this, we examine the condition to have a solution $Z$ in the equations \eqref{Zcond-1} and  \eqref{Zcond-2}.
By setting
\balign{
X&:=Z_{ij}(t) \\
Y&:=Z_{ji}(t) \\
A&:=R_{ij}p_j(t) \\
B&:=R_{ji}p_i(t),
}
the equations \eqref{Zcond-1} and  \eqref{Zcond-2} is expressed as
\balign{
AX^2+BY^2&=F, \lb{Fisher-best-mid1} \\
AX-BY&=A-B, \lb{Fisher-best-mid2}
}
and our task is to find a solution $X$ and $Y$ for given $A$, $B$, and $F$.
Substituting \eref{Fisher-best-mid2} into \eref{Fisher-best-mid1} in order to eliminate $Y$, we obtain the following quadratic equation of $X$:
\eq{
A(A+B)X^2-2A(A-B)X+(A-B)^2-BF=0.
}
The discriminant of this quadratic equation is equal to
\eq{
D=4AB((A+B)F-(A-B)^2).
}
Hence, there exists a real solution of $X$ if and only if $D\geq 0$, which reads
\eq{
F\geq F^*:=\frac{(A-B)^2}{A+B}=\frac{(R_{ij}p_j(t)-R_{ji}p_i(t))^2}{R_{ij}p_j(t)+R_{ji}p_i(t)}.
}
This relation means that the choice of $Z_{ij}$ in \eref{Z-choice} and $F_{ij}$ in \eref{Fij-choice} is the best parameter choice for a tight inequality.
We remark that the above argument also implies that for any $F_{ij}\geq F_{ij}^*=\dot{\Pi}_{ij}$ there always exists a parameter $Z_{ij}$ with which the Fisher information $F(\theta)$ is equal to $\sum_{(i,j)}F_{ij}$.
This fact is employed in the paper of Liu, Gong, and Ueda~\cite{LGU20}, where they show the existence of a complicated form of $Z_{ij}$ with which the Fisher information is exactly equal to the entropy production rate $\dsgm$.
However, if $F_{ij}\neq \dPi_{ij}$, the parameter $Z_{ij}$ is no longer anti-symmetric (i.e., $Z_{ij}=-Z_{ji}$) and thus we cannot set the generalized current $J_d$ to $Z$, which is required to attain the equality of this inequality as we will see in \sref{observable}.

\subsection{Reduction of other TUR-type inequalities}

The master inequality \eqref{master-TUR} (and its stationary version \eqref{master-TUR-ss}) reproduces various known TUR-type inequalities.
We have already seen that the conventional TUR \eqref{TUR} with entropy production is reproduced by the master inequality \eqref{master-TUR-ss} by applying
\eq{
\dot{\Pi}(t)\leq \dsgm(t).
}
A generalized TUR with arbitrary initial states \eqref{GTUR1} is also reproduced by \eref{master-TUR}.

\bigskip

We next reproduce the {\it kinetic uncertainty relation} (KUR) shown by Terlizzi and Baiesi~\cite{TB19}.
They employ not entropy production but activity $A$ to bound the variance, which is defined as
\eq{
A(t):=\sum_{i,j(\neq i)} R_{ij}p_j(t).
}
The activity quantifies the frequency of jumps in the system and thus can be regarded as a measure of time-scale.
The activity frequently appears in nonequilibrium statistical mechanics, from the glassy dynamics~\cite{Gar07, LAW07, BT12} to the characterization of nonequilibrium steady state~\cite{Maes1, Maes2}.
Using the time-integrated activity $\calA:=\int_0^\tau dt A(t)$, Terlizzi and Baiesi showed the following TUR-type inequality in the stationary state named the kinetic uncertainty relation (KUR):
\eqa{
\frac{(\calJ_d(\tau))^2}{\Var(\calJ_d)}\leq \calA.
}{KUR}
Several numerical simulations~\cite{TB19, HS21} imply that KUR becomes a good bound in systems far from equilibrium.
This shows clear contrast to the original TUR, which becomes a good bound in systems close to equilibrium.

The KUR \eqref{KUR} can also be derived from the master inequality \eqref{master-TUR-ss}.
Using $R_{ij}p_j(t)\geq 0$, and $R_{ji}p_i(t)\geq 0$, it is easy to get
\eqa{
\dot{\Pi}(t)=\sum_{i\neq j}\frac{(R_{ij}p_j(t)-R_{ji}p_i(t))^2}{R_{ij}p_j(t)+R_{ji}p_i(t)}\leq \sum_{i\neq j} R_{ij}p_j(t)+R_{ji}p_i(t)=2A(t),
}{Pi-A}
which reproduces the kinetic uncertainty relation \eqref{KUR}.

Derivations of TUR and KUR from the master inequality \eqref{master-TUR-ss} elucidate the difference between when these two inequalities become good bounds.
The inequality \eqref{Pi-sigma} between $\dPi$ and $\dsgm$ becomes tight when $R_{ij}p_j(t)\simeq R_{ji}p_i(t)$ is satisfied, which realizes in systems close to equilibrium.
In particular, $\dPi=\dsgm$ means that this system is in equilibrium.
In contrast, the inequality \eqref{Pi-A} between $\dPi$ and $A$ becomes tight when $R_{ij}p_j(t)\ll R_{ji}p_i(t)$ or $R_{ji}p_i(t)\ll R_{ij}p_j(t)$ is satisfied, which realizes in systems far from equilibrium and the stochastic current flows almost one direction.

\bigskip

The master inequality \eqref{master-TUR} also reproduces the TUR-type inequality for unidirectional transitions shown by Pal, Reuveni, and Rahav~\cite{PRR21}.
A transition $j\to i$ is called unidirectional if this transition occurs with nonzero probability (i.e., $R_{ij}>0$), while its backward transition is prohibited (i.e., $R_{ji}=0$).
In Markov jump processes with some unidirectional transitions, Pal, Reuveni, and Rahav proposed a bound not only with the (partial) entropy production~\cite{SS15} $(R_{ij}p_j(t)-R_{ji}p_i(t))\ln [R_{ij}p_j(t)/R_{ji}p_i(t)]$ but with a quantity $R_{ij}p_j(t)$ on unidirectional edges.
By denoting by $B$ and $U$ the set of edges with bidirectional and unidirectional transitions respectively, their bound reads
\eqa{
\frac{(\tau J_d(\tau))^2}{\Var(\calJ_d)}\leq \frac{\sigma_B+j_U}{2},
}{uni-TUR}
where we defined
\balign{
\sigma_B&:=\int_0^\tau dt \sum_{(i,j)\in B} (R_{ij}p_j(t)-R_{ji}p_i(t))\ln \frac{R_{ij}p_j(t)}{R_{ji}p_i(t)}, \\
j_U&:=\int_0^\tau dt \sum_{i,j: (i,j)\in U}R_{ij}p_j(t).
}

This TUR-type inequality \eqref{uni-TUR} can also be derived from the master inequality \eqref{master-TUR}.
In fact, if a transition $j\to i$ is unidirectional, we have $R_{ji}=0$ and thus $\dot{\Pi}_{ij}$ reduces to
\eq{
\dot{\Pi}_{ij}=\frac{(R_{ij}p_j(t))^2}{R_{ij}p_j(t)}=R_{ij}p_j(t),
}
with which the TUR-type inequality \eqref{uni-TUR} is reproduced.

\section{Optimal observable with generalized empirical measure}\lb {s:observable}

\subsection{Optimal thermodynamic uncertainty relation and achivability of equality}

Even with this tightest parameter $F^*=\dPi$, the equality of the master inequality \eqref{master-TUR-ss} is not achieved in general nonequilibrium systems.
In this section, we extend the class of observables to achieve equality in general nonequilibrium systems.
We here consider only the stationary systems for simplicity.

We extend the class of observables from generalized currents $\calJ_d$ to the sum of generalized currents and generalized empirical measures.
The (conventional) empirical measure counts the staying time at a state, whose ensemble average is nothing but the probability distribution.
The empirical measure appears in various studies in nonequilibrium statistical mechanics from the large deviation theory~\cite{DV75, Shi13, BFG15} to steady-state thermodynamics~\cite{MNW08, MN14}.
The generalized time-integrated empirical measure $\calI_q$ is given by
\balign{
\hcalI_q:=&\sum_i q_i \hcalI_i, \\
\hcalI_i:=&\int_0^\tau dt \delta_{w(t), i},
}
with $w(t)$ the state of the system at time $t$.
We set the possible form of observables as $\calJ_d+\calI_q$.
This type of extension has already been seen in \cite{DS21-2, PRR21}.
Remarkably the rescaling of time-scale given by \eref{Zcond-2} does not change the stationary distribution, which implies
\eq{
\frac{\del}{\del \theta}\langle \hcalI_q\rangle_\theta =0.
}
Substituting $g(\Gamma)=\hcalJ_d+\hcalI_q$ into the generalized Cram\'{e}r-Rao inequality \eqref{CR} and following the derivation of the master inequality \eqref{master-TUR-ss}, we have the optimal TUR-type inequality:
\eqa{
\frac{(\calJ_d)^2}{\Var(\calJ_d+\calI_q)}\leq \frac{\Pi}{2}.
}{best-TUR}
The interesting point of this inequality lies in the fact that the equality in this inequality \eqref{best-TUR} is achieved in general nonequilibrium systems by optimizing over all $J_d$ and $I_q$:
\eq{
\max_{J_d, I_q}\frac{(\calJ_d)^2}{\Var(\calJ_d+\calI_q)}=\frac{\Pi}{2}.
}
With this respect, we call this inequality \eqref{best-TUR} as the {\it optimal} TUR-type inequality in Markov jump processes.

We finally discuss what $\calJ_d$ and $\calI_q$ achieves the equality.
The equality condition of \eref{best-TUR} comes from that of the generalized Cram\'{e}r-Rao inequality, and the generalized Cram\'{e}r-Rao inequality is a special case of the Schwarz inequality.
The Schwarz inequality $(\sum_i a_i^2)(\sum_i b_i^2)\geq (\sum_i a_ib_i)^2$ achieves its equality if and only if $a_1/b_1=a_2/b_2=\cdots=a_n/b_n$ is satisfied.
Thus the equality condition of the generalized Cram\'{e}r-Rao inequality reads
\eq{
g(\Gamma)=\frac{\del}{\del \theta} \ln P_\theta(\Gamma)
}
for any trajectory $\Gamma$.
Recalling the form of transition matrix of $R^\theta$ given in Eqs.~\eqref{Rtheta-1} and \eqref{Rtheta-2}, the above condition is fulfilled with setting
\balign{
d_{ij}&=\frac{R_{ij}\pss_j-R_{ji}\pss_i}{R_{ij}\pss_j+R_{ji}\pss_i}, \lb{TUR-optimal-d} \\
q_j&=-\sum_i \frac{R_{ij}\pss_j-R_{ji}\pss_i}{R_{ij}\pss_j+R_{ji}\pss_i}R_{ij}, \lb{TUR-optimal-q}
}
in $J_d$ and $I_q$.
The above $d_{ij}$ is equal to $Z_{ij}$, which is the first leading order of the perturbation on the transition rate, and $q_j$ is the first leading order of the perturbation on the escape rate.
With this $q_j$, the average contribution of $I_q$ from the state $j$;
\eq{
q_j \pss_j=-\sum_i d_{ij}R_{ij}\pss_j,
}
quantifies the local mean of the current $J_d$ at the state $j$ (with sign inversion).
We note that the above $d_{ij}$ and $q_j$ realize $\langle \hJ_d+\hcalI_q\rangle=0$, and thus the variance of $\calJ_d+\calI_q$ is equal to the second moment of $\calJ_d+\calI_q$, where $I_q$ plays the role to subtract the local mean of $J_d$.

\subsection{Unattainability of equality of the TUR with only a generalized current in nonequilibrium systems}

We shall show that even the master inequality \eqref{master-TUR-ss} can achieve its equality only when the system is in equilibrium.

The argument in the previous subsection directly implies that the master inequality \eqref{master-TUR-ss} achieves its equality if $I_q$ with \eref{TUR-optimal-q} is identically zero, which happens only if
\eqa{
q_j=\sum_i Z_{ij}R_{ij}=\sum_i \frac{R_{ij}\pss_j-R_{ji}\pss_i}{R_{ij}\pss_j+R_{ji}\pss_i}R_{ij}=0
}{TUR-q=0}
is satisfied for all $j$.
If \eref{TUR-q=0} is satisfied, the equality is attained by setting $d$ as \eref{TUR-optimal-d}.

Suppose that both the stationary condition
\eq{
\sum_i J_{ij}=\sum_i (R_{ij}\pss_j-R_{ji}\pss_i)=0
}
and the condition \eqref{TUR-q=0} are satisfied for all $j$.
By definitions of $Z_{ij}$ and $J_{ij}$, it is easy to have
\balign{
J_{ij}&=J_{ij}\frac{2R_{ji}\pss_i+R_{ij}\pss_j-R_{ji}\pss_i}{R_{ij}\pss_j+R_{ji}\pss_i} \nt \\
&=\frac{J_{ij}}{R_{ij}\pss_j+R_{ji}\pss_i}2R_{ij}\pss_j-J_{ij}\frac{R_{ij}\pss_j-R_{ji}\pss_i}{R_{ij}\pss_j+R_{ji}\pss_i} \nt \\
&=2Z_{ij}R_{ij}\pss_j-J_{ij}Z_{ij}.
}
Summing both-hand sides over $i$, we arrive at 
\eq{
0=\sum_i J_{ij}=\sum_i [2Z_{ij}R_{ij}\pss_j-J_{ij}Z_{ij}]=-\sum_i J_{ij}Z_{ij}=-\sum_i \frac{(J_{ij})^2}{R_{ij}\pss_j+R_{ji}\pss_i}.
}
With noting $R_{ij}\pss_j+R_{ji}\pss_i>0$, this relation suggests $J_{ij}=0$ for all edges, which means that this system is in equilibrium.

This observation answers the question of why various TUR-type inequalities do not achieve their equalities in noequilibrium situations.
In fact, we cannot achieve equality as long as we observe only the generalized current, and extending the observables to the generalized empirical measure is crucial for achieving equality.

\section{Relationship to other results}\lb{s:others}

\subsection{TUR for overdamped Langevin systems}

We now consider the (overdamped) Langevin limit of the optimal inequality \eqref{best-TUR}.
We consider a one-dimensional overdamped Langevin system for simplicity.
The dynamics is given by the following Fokker-Planck equation:
\eq{
\frac{\del}{\del t}P(x,t)=-\frac{\del}{\del x} \( \frac{1}{\gamma }F(x)P(x,t)\) +\frac{\del^2}{\del x^2}\( \frac{T}{\gamma }P(x,t)\) .
}
We set $\gamma$ and $T$ as constants for simplicity.
We first discretize the space into a lattice with the lattice interval $\ep$.
The transition rate in this discretized system is given by
\eq{
P_{x\to x\pm \ep}=\frac{1}{\beta \gamma\ep^2}e^{\mp \beta F(x)\ep/2}.
}
We consider $\ep\to 0$ limit of this system and observe what inequality is derived from \eref{best-TUR} in this limit.
The pseudo entropy production rate $\dPi$ in $\ep\to 0$ limit reads
\balign{
\dPi=&\int dx \( \frac12 (\beta F(x)-\frac{d}{dx}\ln \pss (x))\ep \) ^2 \frac{1}{\beta \gamma \ep^2}\pss(x) \nt \\
=&\int dx \frac{\beta \gamma (J^{\rm irr})^2}{\pss(x)} \nt \\
=&\dsgm, 
}
where $J^{\rm irr}$ is the irreversible current:
\eq{
J^{\rm irr}:=\( \frac{1}{\gamma} F(x)-\frac{1}{\beta \gamma}\frac{d}{dx}\ln \pss(x)\) \pss(x).
}
In other words, the inequality $\Pi\leq \sigma$ becomes an equality in the Langevin limit.
We need to distinguish two quantities, the pseudo entropy production and the entropy production, in Markov jump processes, while they collapse into a single quantity in the Langevin systems.

Hence, \eref{best-TUR} reduces to
\eq{
\frac{\calJ_d^2}{{\rm Var}(\calJ_d+\calI_q)} \leq \frac{\sigma}{2}.
}
By setting $I_q$ to the minus of the local mean of the current $J_d$, this inequality reproduces the relative thermodynamic uncertainty relation (RTUR) derived in Dechant and Sasa~\cite{DS21}.

\subsection{TUR for short-time limit and inference of entropy production}

We here consider TUR-type inequalities in the short-time limit for instant quantities, which has been considered in Refs.~\cite{SST16, SS19, Ito18, SFS18, VVH20, PRS17, DS18-PRE, Li19, Ots20}.
One of the advantages of the short-time TURs is that we need not to keep the stationary state unchanged under the time evolution of the $\theta$-system.
Owing to this fact, we can employ the partial entropy production rate~\cite{SS15}
\eqa{
\dsgm_X:=\int_0^\tau dt \sum_{(i,j)\in X} (R_{ij}p_j(t)-R_{ji}p_i(t))\ln \frac{R_{ij}p_j(t)}{R_{ji}p_i(t)}
}{partial-TUR}
instead of the total entropy production $\dsgm$, where $X$ is a set of observed edges.
The partial entropy production is a decomposition of entropy production to each transition, whose sum over all edges recovers the total entropy production.
By restricting the current $J_d$ to the edges in $X$ (i.e., setting $d_{ij}=0$ for $(i,j)\notin X$), we can obtain the TUR for short-time limit:
\eq{
\frac{(J_d)^2}{\Var(J_d)}\leq \frac{\dsgm_X}{2}.
}
Since the $\theta$-modification applies only on the edges in $X$, the probability currents only on the edges in $X$ become $(1+\theta)$-fold, and thus the stationary distribution is no longer $\pss$.
This means that we cannot extend \eref{partial-TUR} to the case of a finite time interval.

\section{Discussion}

We showed that the master inequality with the pseudo entropy production, \eqref{master-TUR} and \eqref{master-TUR-ss}, is the tightest TUR-type inequality for generalized currents.
In addition, we derived the optimal TUR \eqref{best-TUR} by extending observables to both generalized currents and generalized empirical measures, which achieves its equality in general nonequilibrium stationary systems.
This result also reveals why previous TUR-type inequalities are not tight.
We summarize the obtained hierarchical structure in the stationary case:
\eq{
\max_{J_d}\frac{(\calJ_d)^2}{\Var(\calJ_d)}\leq \max_{J_d, I_q}\frac{(\calJ_d)^2}{\Var(\calJ_d+\calI_q)}=\frac{\Pi}{2}\leq \frac{\calF}{2}.
}
The tightness of master inequality \eqref{master-TUR-ss} is manifested in the right inequality.
The left inequality becomes equality only in the equilibrium case, which means the unattainability of the equality in conventional TUR-type inequalities.
In the overdamped Langevin systems, the right inequality becomes equality with $\calF=\sigma$.


Our result suggests the importance of generalized empirical measures in the inference of entropy production.
As we have seen, taking into account generalized empirical measures always improves the bound in nonequilibrium systems.
Another important implication is the importance of the pseudo entropy production $\Pi$.
Since the pseudo entropy production and the entropy production coincide in the Langevin systems, this discrepancy is sometimes overlooked.
This discrepancy may make physics in Markov jump processes much complicated.

\begin{acknowledgements}
The author is grateful to Andreas Dechant, Sosuke Ito, Takahiro Sagawa, and Shin-ichi Sasa for fruitful discussion and helpful comments.
The author is supported by JSPS KAKENHI Grants-in-Aid for Early-Career Scientists Grant Number JP19K14615. 
\end{acknowledgements}

\end{document}